\documentclass{webofc}
\usepackage[varg]{txfonts}   
\begin{document}
\title{The kaonic atoms research program at DA$\Phi$NE: from SIDDHARTA to SIDDHARTA-2}
%
%

\author{\lastname{A. Scordo}\inst{1}\fnsep\thanks{\email{alessandro.scordo@lnf.infn.it}} \and
        \lastname{A. Amirkhani}\inst{2}\fnsep \and
        \lastname{M. Bazzi}\inst{1}\fnsep \and
        \lastname{G. Bellotti}\inst{2}\fnsep \and
        \lastname{C. Berucci}\inst{1,3}\fnsep \and
        \lastname{D. Bosnar}\inst{4}\fnsep \and
        \lastname{A.M. Bragadireanu}\inst{5}\fnsep \and
        \lastname{M. Cargnelli}\inst{3}\fnsep \and
        \lastname{C. Curceanu}\inst{1}\fnsep \and
        \lastname{A. Dawood Butt}\inst{2}\fnsep \and
        \lastname{R. Del Grande}\inst{1}\fnsep \and
        \lastname{L. Fabbietti}\inst{6}\fnsep \and
        \lastname{C. Fiorini}\inst{2}\fnsep \and
        \lastname{F. Ghio}\inst{1}\fnsep \and
        \lastname{C. Guaraldo}\inst{1}\fnsep \and
        \lastname{R.S. Hayano}\inst{7}\fnsep \and
        \lastname{M. Iliescu}\inst{1}\fnsep \and
        \lastname{M. Iwasaki}\inst{7}\fnsep \and
        \lastname{P. Levi Sandri}\inst{1}\fnsep \and
        \lastname{J. Marton}\inst{1,3}\fnsep \and
        \lastname{M. Miliucci}\inst{1}\fnsep \and
        \lastname{P. Moskal}\inst{8}\fnsep \and
        \lastname{D. Pietreanu}\inst{1,5}\fnsep \and
        \lastname{K. Piscicchia}\inst{1,9}\fnsep \and
        \lastname{H. Shi}\inst{1}\fnsep \and
        \lastname{M. Silarski}\inst{8}\fnsep \and
        \lastname{D. Sirghi}\inst{1,5}\fnsep \and
        \lastname{F. Sirghi}\inst{1,5}\fnsep \and
        \lastname{M. Skurzok}\inst{8}\fnsep \and
        \lastname{A. Spallone}\inst{1}\fnsep \and
        \lastname{H. Tatsuno}\inst{10}\fnsep \and
        \lastname{O. Vazquez Doce}\inst{1,6}\fnsep \and
        \lastname{E. Widmann}\inst{3}\fnsep \and
        \lastname{J. Zmeskal}\inst{1,3}\fnsep 
}

\institute{INFN, Laboratori Nazionali di Frascati, Frascati (Roma), Italy
\and    Politecnico di Milano, Dipartimento di Elettronica, Informazione e Bioingegneria and INFN Sezione di Milano, Milano, Italy
\and	Stefan-Meyer-Institut f\"ur Subatomare Physik, Vienna, Austria
\and	Physics Department, University of Zagreb, Zagreb, Croatia
\and	Horia Hulubei National Institute of Physics and Nuclear Engineering (IFIN-HH), Magurele, Romania
\and	Excellence Cluster Universe, Technische Universität München, Garching, Germany
\and	University of Tokyo, Tokyo, Japan
\and	The M. Smoluchowski Institute of Physics, Jagiellonian University, Kraków, Poland
\and	Museo Storico della Fisica e Centro Studi e Ricerche “Enrico Fermi”, Roma, Italy
\and	Lund Univeristy, Lund, Sweden
          }

\abstract{
The interaction of antikaons with nucleons and nuclei in the low-energy regime represents an active research field in hadron physics with still many important open questions. The investigation of light kaonic atoms, in which one electron is replaced by a negatively charged kaon, is a unique tool to provide precise information on this interaction; the energy shift and the broadening of the low-lying states of such atoms, induced by the kaon-nucleus hadronic interaction, can be determined with high precision from the atomic X-ray spectroscopy, and this experimental method provides unique information to understand the low energy kaon-nucleus interaction at the production threshold. The lightest atomic systems, like the kaonic hydrogen and the kaonic deuterium deliver, in a model-independent way, the isospin-dependent kaon-nucleon scattering lengths. The most precise kaonic hydrogen measurement to-date, together with an exploratory measurement of kaonic deuterium, were carried out in 2009 by the SIDDHARTA collaboration at the DA$\Phi$NE electron-positron collider of LNF-INFN, combining the excellent quality kaon beam delivered by the collider with new experimental techniques, as fast and very precise X-ray detectors, like the Silicon Drift Detectors. The SIDDHARTA results triggered new theoretical work, which achieved major progress in the understanding of the low-energy strong interaction with strangeness reflected by the antikaon-nucleon scattering lengths calculated with the antikaon-proton amplitudes constrained by the SIDDHARTA data. The most important open question is the experimental determination of the hadronic energy shift and width of kaonic deuterium; presently, a major upgrade of the setup, SIDDHARTA-2, is being realized to reach this goal. In this paper, the results obtained in 2009 and the proposed SIDDHARTA-2 upgrades are presented.
}
\maketitle
\section{Introduction}
\label{Introduction}

The strong interaction, described by the QCD in the framework of the Standard
Model, is still hiding many mysteries, especially in the low-energy limit, the so
called non-perturbative regime.
Particularly interesting is the strong interaction involving the strange quark which,
belonging to the light quarks but having a mass of about $100\, MeV/c^2$, much
heavier than the few $MeV/c^2$ up and down quark masses, plays a peculiar role.
Since kaons and antikaons are the lowest mass particles containing strange quarks, 
for decades their interaction with nucleons and nuclei in the low-energy regime has been subject
of intensive studies in experiment and theory (for reviews see \cite{intro1,intro2}). 
Effective field theories contain appropriate degrees of
freedom to describe physical phenomena occurring at the nucleon-meson
scale and chiral perturbation theory was extremely successful
in describing systems like pionic atoms; however it is not directly
applicable for the kaonic systems, where non-perturbative coupled-channel 
techniques can be used (\cite{coupled}). These theories are still waiting to be verified by experimental data.
There are two experimental approaches to probe the kaon-nucleus strong interaction, 
both exploited at LNF-INFN. 
One is by studying the scattering and the reaction channels between the kaon and the nucleus, to directly
search for bound states and extract the potential of the strong interaction; 
this is the experimental method followed by the AMADEUS collaboration (\cite{amad,had20,had21}).
The other method consists in the precision X-ray measurement of the shift and the broadening of the
energy levels of kaonic atoms caused by the kaon-nucleus strong interaction. 
This latter one, exploited by the SIDDHARTA and SIDDHARTA-2 collaborations (\cite{sid1,sid2,sid3,sid4,sid5,sid6,sid7}),
is of significant importance, since it is the only method able to provide direct information on the
kaon-nucleus system at threshold.

\subsection{Kaonic atoms}
\label{Introduction-2}

\begin{figure}[htbp]
\centering
\includegraphics[width=10cm]{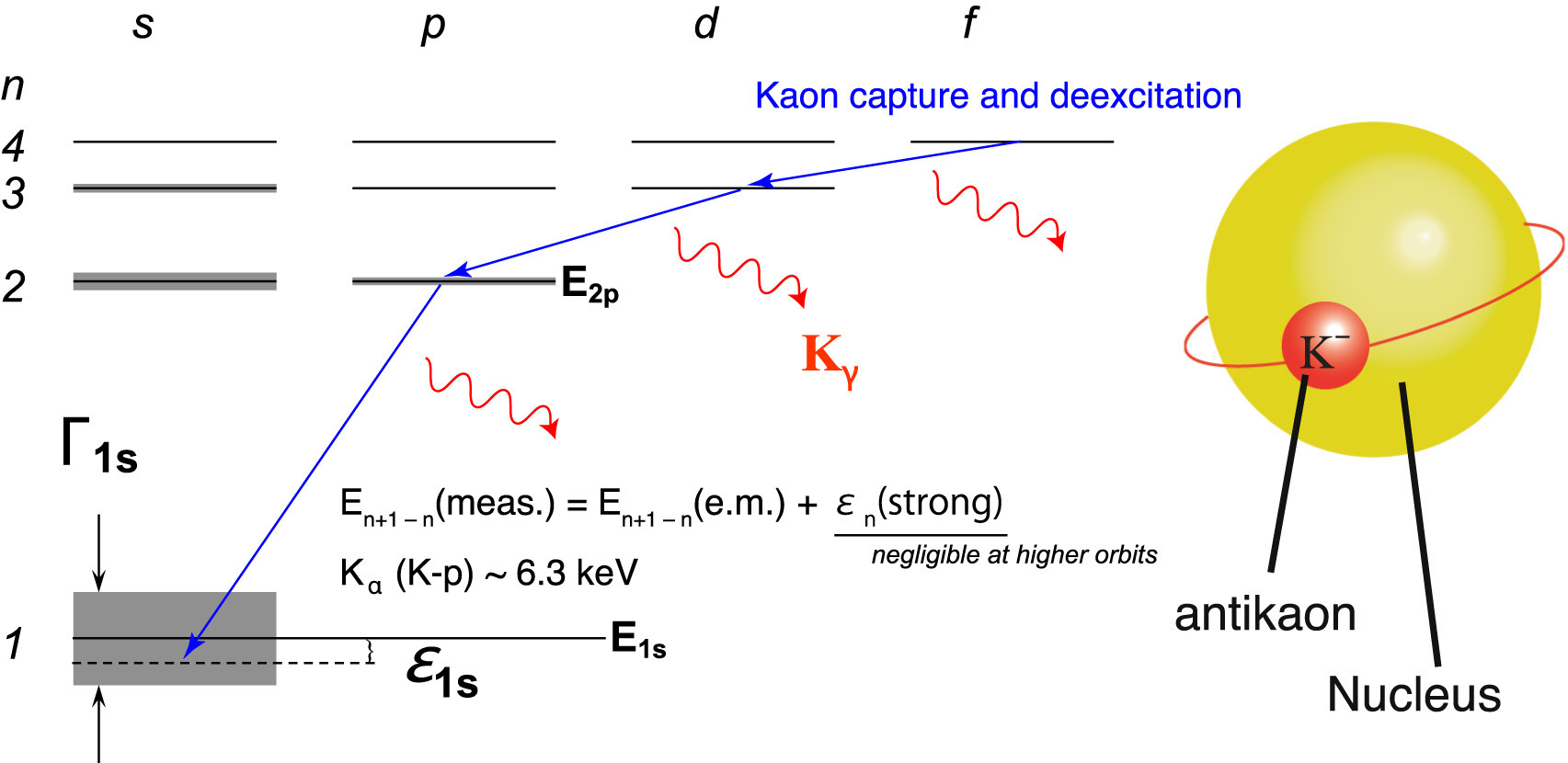}
\caption{After the atomic capture of the kaon a kaonic atom is
formed in a highly excited state and a few kaons will cascade to
ground state. 1s level is shifted and broadened by the strong interaction (\cite{kmass})}
\label{levels}       
\end{figure}

Kaonic atoms are formed when a negatively charged kaon is stopped in a target and replaces one of the electrons in an atom;
due to the much higher $K^-$ mass with respect to the $e^-$ one, this newly formed exotic atom radius is
smaller than the one of the usual atom and the $K^-$ is stopped at a distance from the nucleus corresponding to the $n\simeq25$ excited state.
The subsequent cascade of the antikaon to the ground level occurs via several different processes; 
in particular, the last transitions on the 1s level are radiative and photons are emitted in the X-ray region. 
For light atoms, especially for the hydrogen and the deuterium, a detectable energy
shift from the electromagnetic value of the ground state is expected, as well as
a broadening of the ground state level, caused by nuclear absorption (see fig.\ref{levels}).

By measuring these observables, kaonic atoms offer the unique possibility to determine
the s-wave antikaon-nucleon scattering lengths at vanishing energy,
alternatively to scattering experiments were an extrapolation to zero energy is necessary.
With the advance of the experimental techniques,
both in the accelerator and detector sectors, we are presently able to perform very
high precision measurements, resulting in a deeper and more complete
understanding of the many open questions in the QCD. 
In particular, with the advent of clean kaon beams, as
the one provided by the DA$\Phi$NE collider, and of performant, fast, X-ray detectors
as the Silicon Drift Detectors, the kaonic atoms studies entered the precision era.

\section{SIDDHARTA experiment}
\label{2009}

In 2009, with the SIDDHARTA experiment at DA$\Phi$NE,
the strong interaction induced shift of the ground state of kaonic hydrogen atoms
and the absorption width were measured with the highest accuracy up to now \cite{sid2,sid3}.
Measurements of the 2p shift and width of kaonic helium ( $^4He$ as well as $^3He$) \cite{sid4,sid5,sid6} completed the SIDDHARTA program.
Conscient about the importance of the Kd measurement an attempt was also done in this direction but,
due to the very low yield of the transition, only an upper limit could be exctracted and no values for the
1s level shift and width were delivered \cite{sid2}.

\subsection{Experimental setup}
\label{2009-2}

\begin{figure}[htbp]
\centering
\includegraphics[width=7cm]{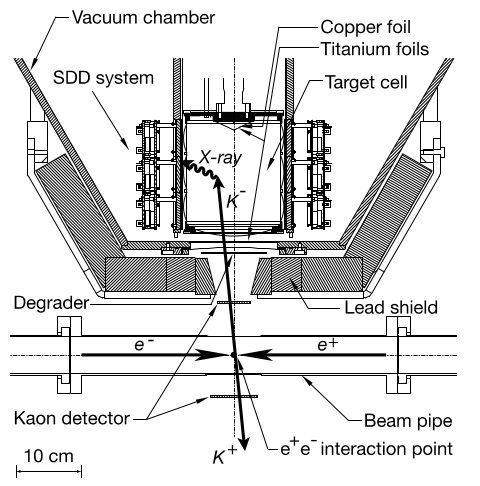}
\caption{A schematic cutaway view of the SIDDHARTA setup at the DA$\Phi$NE interaction point.
The charged kaon pairs are identified with two plastic scintillators, and the $K^-$ induced x-rays
detected by the SDDs are identified from the time correlation to the kaon pair events.}
\label{setup-2009}       
\end{figure}

The SIDDHARTA setup consisted of two main components, the light-weight cryogenic
target cell and a specially developed large area, high resolution X-ray detector system
made of Silicon Drift Detectors (SDDs).
The experiment made use of the $K^+K^-$ pairs coming from the $\Phi$ decays with a 49\% branching
ratio. 
The kaons leaving the interaction point through
the SIDDHARTA beam pipe were degraded in energy and entered the cryogenic gaseous
hydrogen (helium) target placed above the beam pipe, forming a kaonic atom and emitting X-rays
during the $2p\rightarrow1s$ (hydrogen) and $3d\rightarrow2p$ (helium) transitions (see fig. \ref{setup-2009}). 

The $K^+K^-$ pairs were emitted in back-to-back configuration and identified by the kaon detectors,
made of two plastic scintillators placed above and below the interaction point as illustrated in
fig. \ref{setup-2009}. The kaons were distinguished from the minimum ionizing particles using the time of
flight information at the kaon detectors, and the coincidence of the two scintillators defined the
kaon trigger, which marked the timing of the incident kaons. 
A fraction of the negatively charged kaons activating the kaon trigger were successfully stopped inside the
volume of the gaseous target placed about $20\, cm$ above the interaction point to form kaonic
atoms. A working pressure of 0.3 MPa was achieved, which led to a hydrogen gas
density of 2\% of liquid hydrogen density, at a working temperature of 25 K.
Special Silicon Drift Detectors were developed with excellent energy resolution ($FWHM \simeq\, 150\, eV$ @ 6 keV) 
and timing capability in the order of $\mu s$ \cite{sdd} when operated at $140\,K$ temperature. 
Using the X-ray signal from the SDDs in coincidence with the $K^+K^-$ pair, the continuous
machine background, as well as unwanted fluorescence X-rays, could be efficiently suppressed.
The SDDs had a total active area of $144\, cm^2$ , covering about 10\% of the solid angle around
the target cell. The energy calibrations of the SDDs were done every few hours,
using the X-ray tube activated $K_{\alpha}$ lines of Ti ($4.5\, keV$), Mn ($5.5\, keV$) and Cu ($8.0\, keV$) to determine the scale of the energy
spectra. The energy resolution at $6\, keV$ was stable at about $150\, eV$ (FWHM) throughout the
measurement. More details about the configuration and the performance of the detectors can
be found in \cite{sid2,sid3,sid4,sid5,sid6}.
Data were accumulated with gaseous targets of hydrogen ($1.3\,g/l$), 
deuterium ($2.50\,g/l$), helium-3 ($0.96\, g/l$), and helium-4 ($\,1.65 g/l$ and $2.15\, g/l$).

\subsection{Kaonic Helium measurements}
\label{2009-3}

\begin{figure}[htbp]
\centering
\mbox{\includegraphics[width=11.cm]{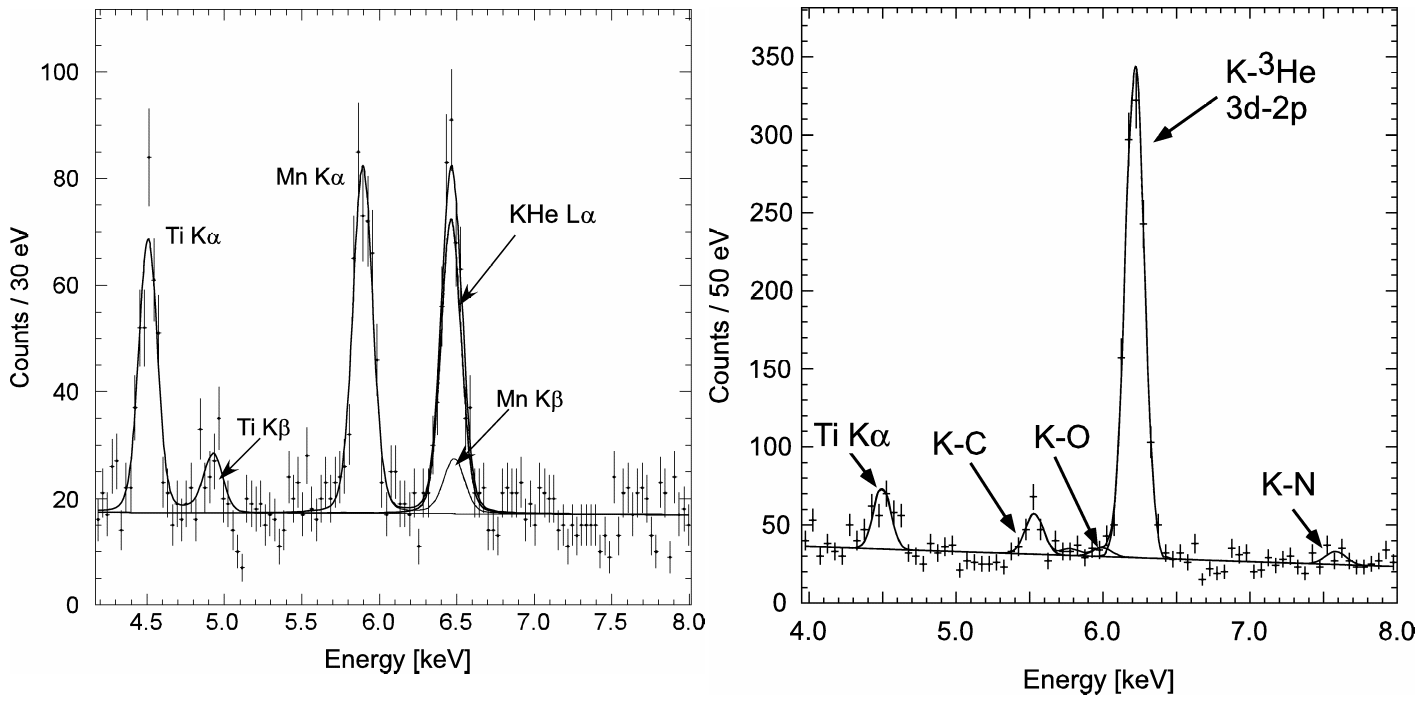}}
\caption{Fitted spectra of the kaonic $^3He$ (right) and $^4He$ (left) X-rays. The $3d\rightarrow2p$
transitions are seen around 6 keV. Together with these
peaks, others smaller are seen, which are the kaonic atom X-ray lines produced by
kaons stopping in the target window made of Kapton, and the Ti and Mn 
lines.}
\label{SPECHEL}
\end{figure}

For what concerns the kaonic helium, its transition to 1s level couldn't be
observed due to the very low yield and to a transition energy out of the SDD dynamic range. 
Before the SIDDHARTA experiment, there existed only four measurements and the situation was rather ambiguous:
three measurements \cite{HEL1,HEL2,HEL3}, performed more than
20 years ago, giving, within few $\sigma$s, results which are more than an order of magnitude
higher with respect to the theoretical predictions and a more recent one,
performed at KEK \cite{HEL4}, which was, instead, compatible with the theoretical predictions (\cite{helpred,helpred2}), but incompatible with the previous experiments.
A conclusive precise measurement on $K^4He$ was then needed in order to solve the "puzzle", together with 
the first measurement of $K^{3}He$, fundamental to obtain valuable information on the $K^-p$ and the $K^-n$ interactions. 
The kaonic helium spectra obtained by the SIDDHARTA experiment are shown in fig. \ref{SPECHEL} \cite{sid4,sid5,sid6}.

In the left picture, the peak seen at 6.2 keV is identified as the $K^3He$ $L_{\alpha}$ line (the $3d\rightarrow2p$
transition). In the right one, the peak seen at 6.4 keV is identified as the $K^4He$ $L_{\alpha}$ line. 
In addition to these lines, other smaller peaks are clearly visible which are the kaonic atom X-ray lines produced by
kaons stopping in the target window made of Kapton, and the Ti and Mn 
lines.
The strong-interaction shifts of the kaonic helium 2p states
were obtained from the difference between the experimentally determined 
values and the QED calculated ones \cite{elcalc1,elcalc2}. The results are:

$$\varepsilon_{2p}(K^3He) = -2 \pm 2 \,(stat) \pm 4 \,(syst) \,eV $$
$$\varepsilon_{2p}(K^4He) = 5 \pm 5 \,(stat) \pm 4 \,(syst) \,eV $$

\noindent Thus a shift compatible with 0 eV of the 2p level from experiment is established, which is in agreement with
the theoretical estimations and, within the errors, with the
results reported by the E570 \cite{HEL4} collaboration, while is not in agreement with the previous measurements (\cite{HEL1,HEL2,HEL3}). 
This is probably due to the fact that, in those experiments, the helium target was a liquid one in which, as indeed happened also in the KEK experiment (\cite{HEL4}), a big Compton effect is present and has to be taken into account in the analysis procedure.

\subsection{Kaonic Hydrogen and Kaonic Deuterium measurement}
\label{2009-4}

The lightest kaonic atom is the $K^-p$ atom in which the principal electromagnetic interaction is 
accompanied by the strong interaction of the kaon with the proton which is measurable by X-ray spectroscopy of the
radiative transitions from the np states (2p, 3p, ...) to the 1s ground state (K transitions).
The $K^-p$ scattering length $a_{K^-p}$ can be obtain from the equation 
$$\varepsilon_{1s}+\frac{i}{2}\Gamma_{1s}=2\alpha^3\mu_c^2a_{K^-p}(1-2\alpha\mu_c(ln\alpha-1)a_{K^-p})$$ 

\noindent being $\varepsilon_{1s}$ and $\Gamma_{1s}$ the shift and width of the transition to the 1s level, 
(for more details on the various terms see \cite{deser}) and it is related to the isospin dependent scattering lengths by 

$$a_{K^-}p=\frac{1}{2}(a_0+a_1)$$

\begin{figure}[htbp]
\centering
\mbox{\includegraphics[width=6.cm]{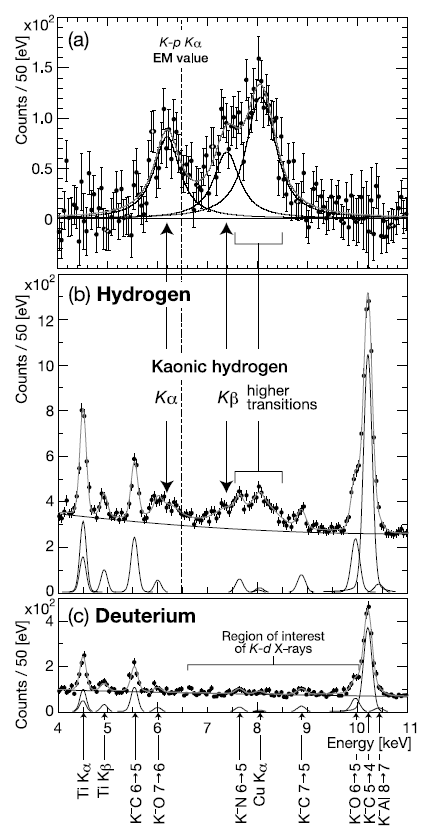}}
\caption{A global simultaneous fit result of the X-ray energy spectra of hydrogen
and deuterium data. (a) Residuals of the measured kaonic-hydrogen X-ray spectrum
after subtraction of the fitted background, clearly displaying the kaonic-hydrogen
K -series transitions. The fit components of the K-p transitions are also shown,
where the sum of the function is drawn for the higher transitions (greater than $K_{\beta}$).
(b), (c) Measured energy spectra with the fit lines. Fit components of the background
X-ray lines and a continuous background are also shown. The dot-dashed
vertical line indicates the EM value of the kaonic-hydrogen $K_{\alpha}$ energy. (Note that
the fluorescence $K_{\alpha}$ line consists of $K_{\alpha 1}$ and $K_{\alpha 2}$ lines, both of which are shown.)}
\label{SPECHYD}
\end{figure}

\noindent Historically there were several measurements of the strong-interaction shift $\varepsilon_{1s}$ and width $\Gamma_{1s}$ of kaonic hydrogen (\cite{KHPuz1,KHPuz2,KHPuz3,KHPuz4,KHPuz5}).
In the 1970s and the 1980s three groups (\cite{KHPuz1,KHPuz2,KHPuz3}) reported a measured
attractive shift (positive $\varepsilon_{1s}$ ), while the information extracted from the analyses of the low energy KN data  (\cite{KHPuz6,KHPuz7,KHPuz8}) shows a repulsive shift (negative $\varepsilon_{1s}$). This
contradiction has been known as the "kaonic hydrogen puzzle".
In 1997, the first distinct peaks of the kaonic-hydrogen X-rays were observed by the KEK-PS
E228 group \cite{KHPuz4} with a significant improvement in the signal-to-background ratio by the use of
a gaseous hydrogen target, where previous experiments had employed liquid hydrogen. It was
crucial to use a low-density target, namely a gaseous target, because the X-ray yields quickly
decrease towards higher density due to the Stark mixing effect. The observed repulsive shift
was consistent in sign with the analysis of the low energy KN scattering data, resolving the
long-standing discrepancy.
More recent values reported by the DEAR group in 2005 \cite{KHPuz5}, with substantially reduced
errors, firmly established the repulsive shift obtained in the previous E228
experiment.
The latest kaonic hydrogen and deuterium X-ray energy spectra, obtained by the SIDDHARTA experiment, are shown in fig. \ref{SPECHYD} \cite{sid2}. 

The K-series X-rays of kaonic hydrogen were clearly 
observed while those for kaonic deuterium were not visible. This
appears to be consistent with the theoretical expectation of lower
X-ray yield and greater transition width for deuterium (\cite{deuprev}) than for kaonic hydrogen. However, the kaonic deuterium spectrum can be used to characterize the background. 
The vertical dot-dashed line in fig.\ref{SPECHYD} indicates the X-ray energy
of kaonic hydrogen $K_{\alpha}$ calculated using only the electromagnetic
interaction (EM). Comparing the kaonic hydrogen $K_{\alpha}$ measured peak and the
EM value, a repulsive shift of the kaonic hydrogen
1s energy level is easily seen.
Many other lines from kaonic atom X-rays were detected in both spectra as indicated with arrows
in the figure. These kaonic atom lines result from high n X-ray
transitions of kaons stopped in the target cell wall made of kapton
($C_{22}H_{10}O_5N_2$) and its support frames made of aluminium. There
are also characteristic X-rays from titanium and copper foils installed
for X-ray energy calibration.
A global simultaneous fit of the hydrogen and
deuterium spectra has been performed, whose results are shown in fig. \ref{SPECHYD} (b) and (c).
The kaonic hydrogen lines were represented
by Lorentz functions convoluted with the detector response function,
where the Lorentz width corresponds to the strong interaction
broadening of the 1s state. The region of interest of Kd X-rays is illustrated in fig. \ref{SPECHYD} (c).
The 1s level shift $\varepsilon_{1s}$ and width $\Gamma_{1s}$ of kaonic hydrogen
were determined to be \cite{sid2}: 

$$ \varepsilon_{1s} = -283 \pm 36 \,(stat) \pm 6 \,(syst) \,eV$$
$$ \Gamma_{1s} = 541 \pm 89 \,(stat) \pm 22 \, (syst) \,eV$$

The quoted systematic error is a quadratic summation
of the contributions from the systematic errors on the SDD
gain shift, the SDD response function, the ADC linearity, the low energy
tail of the kaonic hydrogen higher transitions, the energy
resolution, and the procedural dependence shown by independent
analysis.
This measurement represents the best precision available untill today, and it has been used to 
set constraints on the calculated real and imaginary part of the $K^-p$ amplitude (\cite{constr}).

\section{SIDDHARTA-2 experiment}
\label{2019}

The case of kaonic deuterium is more challenging than kaonic hydrogen mainly due to the larger
widths of the K lines and the lower X-ray yield expected. 
Experimentally the case of kaonic deuterium is still open; the SIDDHARTA experiment measured the X-ray
spectrum with a pure deuterium target but, due to the limited statistics and high background,
the determination of $\varepsilon_{1s}$ and $\Gamma_{1s}$ was impossible. An upper limit for the X-ray yield of the K lines
could be extracted from the data: total yield < 0.0143, $K_{\alpha}$ yield < 0.0039 \cite{sid7}.

Since the kaonic deuterium X‐ray measurement represents the most important experimental
information missing in the field of the low‐energy antikaon‐nucleon interactions today,
a new experiment (SIDDHARTA-2) is planned, based on a much improved apparatus.

\subsection{Experimental setup upgrade}
\label{2019-2}

Thanks to the knowledge acquired in 2009 with the SIDDHARTA experiment, a new version
of the experimental apparatus, aiming to increase the signal over
background ratio (S/B) by a factor about 20 allowing the kaonic deuterium measurement, 
has been developed by the SIDDHARTA-2 collaboration.
The major upgrades introduced to fulfill the kaonic deuterium measurement goal, are shown in fig. \ref{setup-2019} and listed below (\cite{mihail}):

\begin{figure}[htbp]
\centering
\mbox{\includegraphics[width=8.cm]{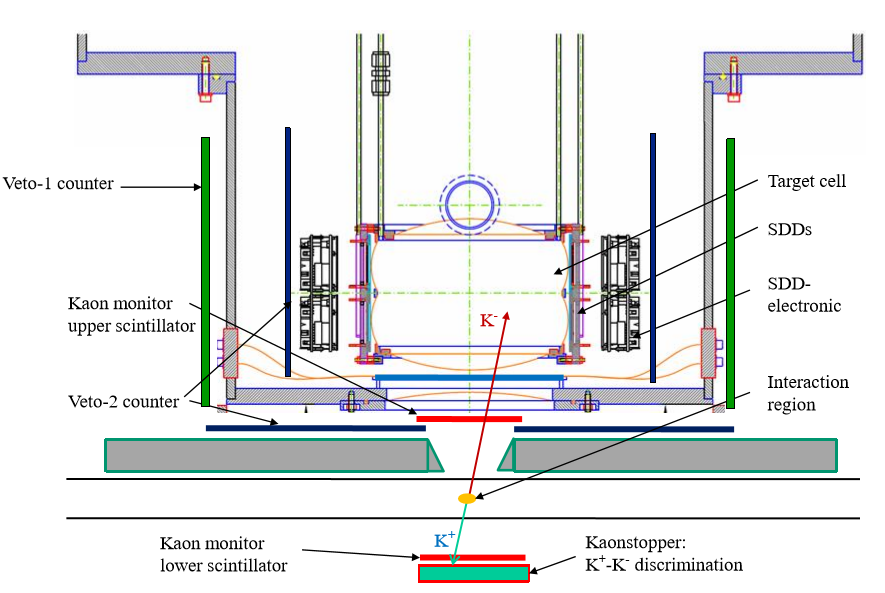}}
\caption{The SIDDHARTA-2 layout. The elements of the apparatus are indicated in the figure.}
\label{setup-2019}
\end{figure}

\begin{itemize}

\item Larger area and faster SDD detector array; the solution to improve the SDD time resolution consists in
the reduction of the single element size (from $10x10$ to $8x8\, mm^2$) and the replacement of the integrated J-FET
(thermally limited), by a newly-developed amplifier on the ceramics, able to operate at very low
temperatures (below 50 K). The shorter path and the higher carrier mobility consent a faster charge
packet drift to the anode and therefore, a reduced time window (350 ns instead of 900 ns) for each
trigger, with consequent a suppression of the asynchronous background. The new SDD detectors are 
produced by Fondazione Bruno Kessler (FBK) in Trento, Italy.

\item A veto detector to measure the prompt time of the secondaries from $K^-$ absorption on nuclei.
The system consists in scintillators surrounding the
vacuum chamber, read at both ends by PMs coupled to mirrors and light-guides (to cope with the
narrow space between the setup and the shielding against beam background) \cite{lshape}.

\item A new cryogenic target in reinforced kapton (13 cm diameter, 7 cm height), operating few hundred
mK above the liquid point (~25 K) at a pressure of 4 bar (5\% LHD), for more efficient kaon stopping.

\item A veto system, consisting in scintillators read by SiPMs, placed behind each SDD array, to reject
the hadronic background coming from border hits of Minimum Ionizing Particles (MIPs), depositing energy in the X-ray range.

\item A kaon trigger with geometric acceptance optimized to match the kaon gas stopping distribution.

\item An improved X-ray calibration system, providing low-rate in-situ calibration, as well as high rate
calibration between physics runs, to compensate very small fluctuations in each single SDD response.

\item Mechanical and cryogenic improvements of the vacuum chamber, necessary to add more cooling
power to the SDDs and to the cryogenic target.

\end{itemize}

\subsection{Kaonic deuterium expected measurement}
\label{2019-3}

\begin{figure}[htbp]
\centering
\mbox{\includegraphics[width=8.cm]{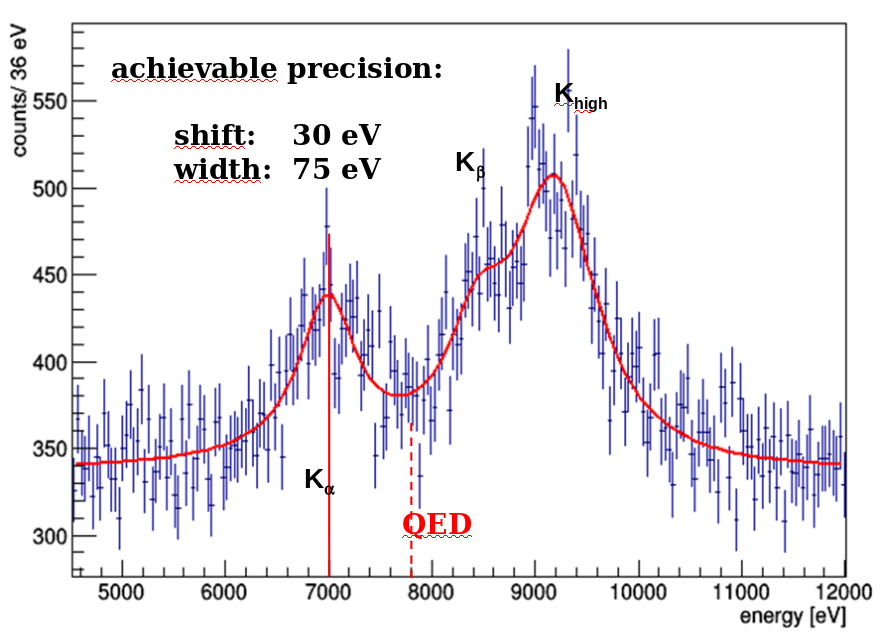}}
\caption{The simulated $K^-d$ spectrum for SIDDHARTA-2 for $800\,pb^{-1}$ integrated luminosity (the $K_{\alpha}$ line is at 7 keV, 
while from 8 to 10 keV there is the K-complex.}
\label{mcdeut}
\end{figure}

All the above described items were optimized by GEANT4 simulation, considered reliable after having
reproduced, in the same framework, the SIDDHARTA results, both in terms of signal and background,
to 7\% accuracy. 
Using theoretical values as inputs to the Monte Carlo simulation, the expected spectrum of the $K^-d$ transitions, 
for an acquired luminosity of $800\, pb^{-1}$ and assuming a yield of 0.1\%, is shown in fig. \ref{mcdeut}. 

\noindent The fit of the simulated spectrum shows how the $K^-d$ $2p\rightarrow1s$ transition, influenced by the strong interaction, 
can be determined with a precision of 30 eV for the shift and 70 eV for the width.

\section{Conclusions}
\label{Conclusions}

The SIDDHARTA experiment at the DA$\Phi$NE electron-positron collider measured the strong interaction induced 
shift and width of kaonic hydrogen and helium transitions with unprecedented accuracy. 
Important implications for the theory of the strong interaction with strangeness in
the low energy regime were provided by the SIDDHARTA constraints. 
In addition, SIDDHARTA measured for the first time ever the kaonic helium 3 transitions to the 2p level and
the first, similar, measurement, of kaonic helium 4 with a gaseous target. 
Still open is the challenging case of kaonic deuterium which will be attacked by the follow-up experiment SIDDHARTA2, aiming at the determination 
of the hadronic shift and width to enable the extraction of the antikaon-nucleon isospin dependent scattering lengths $a_0$ and $a_1$.
The SIDDHARTA-2 collaboration is preparing a series of modifications and upgrades of the apparatus and
experimental configurations, aiming to measure the kaonic deuterium x-rays with a precision
compatible with that achieved in the hydrogen target measurement starting with 2018, when the apparatus will be installed on the DA$\Phi$NE collider.

\section*{Acknowledgments}
\label{Acknowledgments}

We thank C. Capoccia, G. Corradi, B. Dulach, and  D. Tagnani  from  LNF-INFN; 
and  H. Schneider, L. Stohwasser, and D. St\"ukler from Stefan-Meyer-Institut, for their fundamental
contribution in designing and building the SIDDHARTA setup.
We thank as well the DA$\Phi$NE staff for the excellent working conditions and permanent support.
Part of this work was supported by the Austrian  Science  Fund  (FWF)  (P24756-N20);  Austrian  Federal  Ministry  of  Science
and  Research  BMBWK  650962/0001  VI/2/2009;  Croatian  Science  Foundation  under  Project
No.  1680; the Grant-in-Aid for Specially Promoted Research (20002003), MEXT, Japan; 
Ministero degli Affari Esteri e della Cooperazione Internazionale, Direzione Generale per la Promozione del Sistema Paese (MAECI), 
Strange Matter project; the Polish National Science Center (NCN) through grant No. UMO-2016/21/D/ST2/01155.

\end{document}